# Gravitational eigenstates in weak gravity I: dipole decay rates of charged particles

**A D Ernest**

Faculty of Science, Charles Sturt University, Wagga Wagga, 2678, Australia

Email: aernest@csu.edu.au

**Abstract**

The experimental demonstration that neutrons can reside in gravitational quantum stationary states formed in the gravitational field of the Earth indicates a need to examine in more detail the general theoretical properties of gravitational eigenstates. Despite the almost universal study of quantum theory applied to atomic and molecular states very little work has been done to investigate the properties of the hypothetical stationary states that should exist in similar types of gravitational central potential wells, particularly those with large quantum numbers. In this first of a series of papers, we attempt to address this shortfall by developing analytic, non-integral expressions for the electromagnetic dipole state-to-state transition rates of charged particles for any given initial and final gravitational quantum states. The expressions are non-relativistic and hence valid provided the eigenstate wavefunctions do not extend significantly into regions of strong gravity. The formulae may be used to obtain tractable approximations to the transition rates that can be used to give general trends associated with certain types of transitions. Surprisingly, we find that some of the high angular momentum eigenstates have extremely long lifetimes and a resulting stability that belies the multitude of channels available for state decay.

## 1. Introduction

The bound stationary states of electrons in atoms have been studied extensively for many years but little has been done to investigate the properties of neutral or charged particles occupying the hypothetical stationary states in gravitational potential wells. This is important because there are no strong scientific grounds to deny the existence of gravitationally bound quantum eigenstates (including macroscopic ones) and also because recent experimental work by Nesvizhevsky *et al* [1, 2] has demonstrated the physical reality of such gravitational quantum stationary states in the Earth's gravitational field. It is interesting therefore to speculate on whether it might be theoretically possible for relatively pure stationary gravitational eigenstates to exist naturally elsewhere in the universe. If such states were to exist, it would clearly be important to have theoretical information on their expected properties.

This paper aims to set up a simple quantum model using a central gravitational point potential, derive its set of hypothetical gravitational eigenstates and begin an analysis of the properties and interactions of some of these states. Specifically we develop formulae for the form of the dipole-allowed state-to-state transition rates in the general case of charged particles occupying stationary states with arbitrarily large quantum numbers. Although mathematically the eigenstates of such a system are theoretically analogous to those of an equivalently bound electrical system such as the hydrogen atom, differences in the sizes of the quantum numbers and scales involved, combined with the unusual properties of the eigenfunctions themselves, result in some of the gravitational eigenstates having unusual and unexpected properties. Deriving these properties and understanding reasons for the differences between the atomic and gravitational cases is often not trivial. The long term aim is to determine under what conditions long-lived, relatively stable, gravitational eigenstates can exist, and further, to make predictions about the properties of possible (necessarily large scale) hypothetical structures composed of an ensemble of such states. If such relatively pure gravitational eigenstates do exist in the universe then it is possible that they could have significant consequences for Astronomy [3, 4].

The simplest realistic gravitational system consists of a symmetric potential produced by a sufficiently large mass *M* so as to have well bound eigenstates occupied by particles of atomic mass scales. We therefore consider the gravitational quantization of a small particle mass $m_p$ $(<< M)$ in the point potential field provided by the large, electrically neutral, mass *M*. To simplify matters further we consider the states whose radial extents are limited and have appreciable amplitude only over regions which are sufficiently distant from the centre of *M* so that the gravitational field is weak enough to ignore relativistic effects. Under these conditions the Schrödinger equation is trivially analogous to that for the hydrogen atom and may be written as

$$-\frac{\hbar^2}{2\mu}\nabla^2\psi - \frac{Gm_pM}{r}\psi = i\hbar\frac{\partial\psi}{\partial t} \tag{1}$$

where $\mu$ is the reduced mass and the other symbols have their normal meanings.

The solutions are likewise analogous to those of the hydrogen atom and may be immediately written down, the eigenvalues $E_n$ being

$$E_n = -\frac{\mu G^2 m_p{}^2 M^2}{2\hbar^2 n^2} \qquad (2)$$

With the introduction of the parameter $b_0 = \dfrac{\hbar^2}{G \mu m_p M}$, the corresponding eigenfunctions $u_n(\mathbf{r},t)$ are

$$u_{n,l,m}(\mathbf{r},t) = R_{n,l}(r) Y_{l,m}(\theta,\phi) \qquad (3)$$

where $Y_{l,m}(\theta,\phi)$ are the normalized spherical harmonics and

$$R_{n,l}(r) = N_{nl}\left(\frac{2r}{nb_0}\right)^l \exp\left(-\frac{r}{nb_0}\right) L_{n-l-1}^{2l+1}\left(\frac{2r}{nb_0}\right) \qquad (4)$$

In equation (4) $N_{nl} = \left\{ \left(\dfrac{2}{nb_0}\right)^3 \dfrac{(n-l-1)!}{2n(n+l)!} \right\}^{\frac{1}{2}}$ is a normalizing constant and

$$L_{n-l-1}^{2l+1}\left(\frac{2r}{nb_0}\right) = (n+l)! \sum_{k=0}^{n-l-1} \frac{(-1)^{k+2l}\left(\dfrac{2r}{nb_0}\right)^k}{(n-l-1-k)!(2l+1+k)!k!} \qquad (5)$$

are the generalized Laguerre polynomials in their standard form.

In most traditional situations where a quantum approach would be anticipated, the resulting eigenstate binding energies are unrealistically small because the gravitational force is relatively very weak. For example the binding energies $E_n$ of two neutrons (ignoring spin) is much smaller (e.g. $E_1 \sim 10^{-69}$ eV) than the magnitude of typical random background field fluctuations (e.g. $E \sim 10^{-4}$ eV for cosmic microwave background radiation). The binding energy increases with larger masses, but for small quantum numbers the physical extent of the quantum probability density distribution can become much smaller than the corresponding quantum probability distributions of the individual isolated masses themselves making up the structure ($M \sim m \sim 10^{-13}$ kg implies an $n = 1$ eigenstate 'size' of $\sim 10^{-19}$ m compared to a typical 'size' for the individual $10^{-13}$ kg masses of $\sim 10^{-5}$ m for a mass density $\sim 1$ kg m$^{-3}$). The two particle gravity-only Schrodinger equation is inapplicable in this situation since the interactive effects of the electron clouds in the two masses would then dominate. We therefore consider the theoretical description and properties of very large, possibly macroscopic, stationary, well bound states that correspond to large quantum numbers in the (relatively) weak regions of deep gravitational wells.

The notion of large quantum states and macroscopic quantum phenomena is not new. Bose-Einstein condensates, Superconducting Quantum Interference Devices, and the many experiments involving quantum connectedness all demonstrate existence of macroscopic quantum phenomena [5-8]. Of course there are understandable reasons for certain quantum phenomena, such as interference effects, not occurring on large scales. In the case of quantum interference, for example, background electromagnetic or gravitational field fluctuations will generally produce interactions that result in decoherence on macroscopic scales. However these only introduce randomly phase shifts that occur independently to the individual members of a quantum connected pair of virtual states [9] thereby destroying the quantum coherence and the subsequent interference. There is no experimental evidence that this type of macroscopic decoherence should result in the indiscriminate rejection of all macroscopic quantum phenomenon.

Not only will we be concerned with high-$M$ and high-$n$ values, but also the relatively high angular momentum states (high-$l$ relative to $n$) because for these states the radial extent of the probability density 'shell' can be relatively limited (and conversely, low-$l$ relative to $n$ implies large radial eigenfunction spread). When the central mass is small, then, as the $n$ value increases, the binding energy becomes too weak for effective binding. This occurs long before the effective position of a high-$l$ eigenstate is able to outstrip the physical size of the central mass $M$ in question, but can be easily achieved if the central mass is large enough. The lowest central mass limit at which this phenomenon becomes feasible depends on the assumed density of the central mass and on the critical value that one adopts as an 'effective' binding energy. However, for realistic mass densities of $10^0$ - $10^2$ kg m$^{-3}$, and a lower limit binding energy of around 1 eV (i.e. $E_n >>$ average background CMB photon energy) the required central mass $M$ is of the order of $10^{30}$ kg. This explains the difficulty in producing or observing gravitational eigenstates on laboratory or microscopic scales. It is of interest to note that the binding energies of the neutrons observed in the work of Nesvizhevsky *et al* (quantized in an asymmetric wedge potential of a 'hard' mirror and the Earth's ($M < 10^{25}$ kg) gravitational field rather than a central spherically symmetric one) had binding energies of only 1.4 to 4.1 peV for $n = 1$ to 4 yet still demonstrated relatively large 'sizes' by traditional quantum standards (~0.05 mm for $n = 4$) [2].

The solutions to (1) are not trivially dealt with by analogy with those of the hydrogenic wavefunctions because of the necessarily large quantum numbers involved in the formation of energetically realistic gravitational eigenstates. Any interaction properties that depend on the relative energy level spacing and the overlap integrals of states involved can be drastically different from those properties resulting from low-$n$ quantum state overlap integrals. Large quantum number wavefunctions of electrically bound electrons in hydrogen need to be rarely, if ever, considered in the atomic case because their binding energies are too small to remain stable and bound for any reasonable length of time. Describing and dealing with states having high-$n$, $l$, $m$ values is therefore a primary goal of the present work. When $n$ is large the complexity of the individual states and the number of available eigenstate transitions makes the general study of the intrinsic properties and interaction rates of any large array of gravitational eigenstates (referred to from now on as an 'eigenstructure') extremely difficult.

We introduce a schematic (figure 1) showing the traditional quantum parameters $n$, $l$ and $m$. As with the atomic case, the total angular momentum parameter $l$ and its $z$-projection sub-levels $m$, range from 0 to $n-1$ and $-l$ to $l$ respectively. Vertical lines represent states of constant $l$ and horizontal lines states of constant $n$. Each solid circle in figure 1 represents the $2l+1$ gravitational $z$-projection sublevels. It is convenient to introduce a quantity $p$ ($\equiv n-l$) that then ranges from 1 to $n$. The reason for this is that it will turn out that the states of interest here will have $p$ values that are small relative to $n$ and formulae for these eigenfunctions are more conveniently written in terms of $p$. The diagonal dotted lines of figure 1 represent lines of constant $p$ and all states lying on one of these diagonal lines all share a common $p$ value, beginning with $p=1$ on the leftmost diagonal. The parameter $p$ reflects the general spatial profile of the radial eigenfunction component: $p=1$ corresponds to a single-peaked, radial eigenfunction component with no zeros, $p=2$ a two-peaked, one-zero function, $p=3$ a three-peaked, two-zero function and so on.

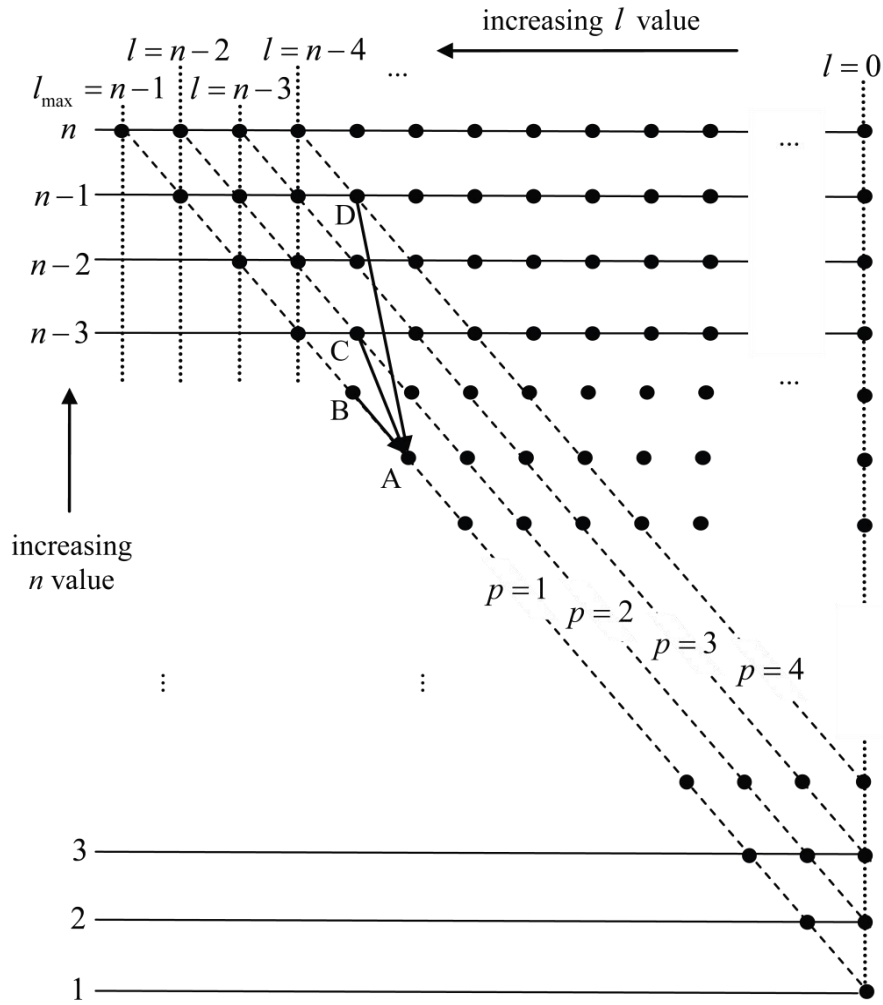

**Figure 1.** Schematic representing the high $n$, $l$, $m$-valued stationary states, drawn to emphasize the parameter $p \equiv n - l$. Each solid circle on the diagram represents $2l + 1$ $z$-projection substates.

To aid the physical understanding of the behaviour of large quantum-valued eigenstates we end this introduction by giving examples of the eigenfunction probability functions $u_n^*(\mathbf{r},t) u_n(\mathbf{r},t)$ of typical gravitational systems for instructive $n$, $l$, $m$ values (essentially a revision of traditional hydrogenic systems) and note some relevant features. The spherical harmonic components $Y_{l,m}(\theta,\phi)$ of the eigenfunctions are explicitly given by

$$Y_{l,m}(\theta,\phi) = \frac{(-1)^m}{2^l\, l!} \sqrt{\frac{2l+1}{4\pi}\frac{(l-m)!}{(l+m)!}} \left(\sin\theta\right)^m e^{im\phi} \sum_{k=0}^{l} \frac{(-1)^{l-k}\, l!}{k!\,(l-k)!} \frac{(2k)!\left(\cos(\theta)\right)^{2k-l-m}}{(2k-l-m)!}$$

$$(6)$$

The azimuthal dependence on $\phi$ involves only functions of the form $e^{\pm im\phi}$ and as with the atomic case, gravitational eigenfunctions result in probability density functions that are completely delocalized in the $\phi$ direction. Hence, one simple representation of the density functions of the stationary states presents them as three dimensional revolutions around the $\theta = 0$ axis of two dimensional $r$, $\theta$ polar coordinate plots shown here by defining a radial-polar probability surface density associated with the probability of finding a particle between $(r, r\theta)$ and

$(r+dr, r(\theta+d\theta))$ as $\frac{d^2P}{dr\,d\theta} = \int_0^{2\pi} \frac{d^3P}{dV}\,d\phi$ (where $dV \equiv$ infinitesimal volume

element). Examples of the radial-polar probability density dependence on $n$, $l$, $m$ may then be drawn (taking an arbitrary scale parameter $b_0$=1) as in figure 2. They are shown here to emphasize that (a) the number of peaks in the density function in the radial and polar directions is given by $p$ and $(l - m + 1)$ respectively and (b) the radial 'thickness' or extent of the state probability density reduces with decreasing $p$. We also note the other well known hydrogenic eigenfunction properties that the average radial position of the density function increases as $n$ increases and also that, for a given normalized set of states with the same $n$, $l$ values, the summed probability density over all possible $m$ values gives a symmetric distribution whose integral over all space is $2l + 1$. For arrays covering sufficiently large $n$ and $l$ values, the sum over a sparsely populated set of $m$ states will also show spherical symmetry provided that the occupancy is randomly distributed. The degree of localization in the $r$ and $\theta$ directions varies with values of $m$ and $l$ but, predictably, none show the degree of localization exhibited by traditional orbiting particles. It is clear that the quantum representation of any spatially-localized orbiting particle in terms of the eigenstate array will necessarily involve a summation over many eigenstates and that the tighter the localization constraint the more expansive is the required set of eigenfunctions.

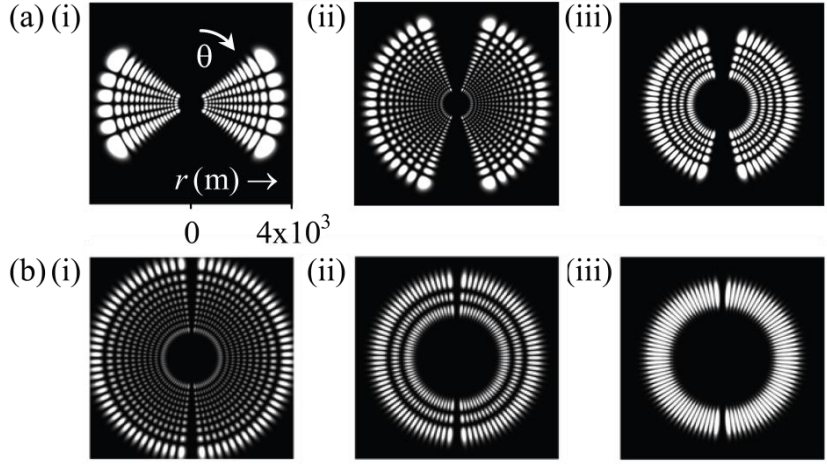

**Figure 2.**
(a) Peaks in the radial-polar density profile reflect the values of $p$
and $l$-$m$+1. $b_0 = 1$, $n = 45$, (i) $p = 15$, $l$-$m$+1 = 6; (ii) $p = 15$, $l$-$m$+1
= 21; (iii) $p = 5$, $l$-$m$+1 = 30;
(b) Radial 'thickness' of density function decreases with
decreasing $p$.
$b_0 = 1$, $n = 50$, $m = 3$, (i) $p = 10$; (ii) $p = 3$; (iii) $p = 1$;

For large quantum numbers calculation of the various overlap integrals for
spontaneous dipole decay is particularly difficult and it is useful to derive non-
integral formulae for these which in turn lead to more manageable formulae for
calculating state-to-state transition rates and ultimately state lifetimes. This is
done in section 2. Despite the more manageable forms of the non-integral
formulae for dipole decay, it becomes necessary to introduce approximations for
specific cases. One case of relevance to the present work is that where one or
more of the states has a $p$ quantum value of 1 (see figure 1). Section 3 deals with
the development of two useful approximate formulae which enable calculations
for these types of transitions. Lastly we conclude with a discussion of the
findings of this paper, their relevance to speculations on the existence of
naturally occurring gravitational eigenstates in the universe, and the need for the
development of further approximations that will be the subject of another
companion paper.

## 2. Exact non-integral state-to-state transition rate formulae

The interactive properties of particles in gravitational eigenstates will be determined by an overlap integral involving a relevant interaction potential. It is not possible in the present treatment to cover all possible types of interaction potentials. In the atomic case the most basic questions concern the stability and decay of the electronic states. We therefore restrict attention initially in this paper to the calculation of decay rates and the intrinsic stability of the states. Subsequently it will be shown in later papers that the mechanisms determining some of the important properties of certain of the states discussed here are also applicable to more general interactions because of intrinsic characteristics of the eigenstates themselves, and are not greatly dependent on the form of the interaction potential that is part of the overlap integral. State-to-state transition rates and state decay rates may therefore enable the estimation of rates for other types of interactions as well.

The central mass is assumed neutral so that eigenstate radiative decay takes place via emission of either gravitational radiation or, if the eigenstate particle is charged, a combination of gravitational and electromagnetic radiation. It has been shown [4] that for the typical conditions considered here radiative decay through gravitational radiation is insignificant compared to that produced through electromagnetic dipole decay. We therefore derive formulae for the state-to-state electromagnetic dipole decay rates that would be applicable, for example, to the two most stable particles, the electron and proton in various eigenstate configurations.

If the eigenstructure is sufficiently densely occupied, then fields of the eigenstate particles themselves will also contribute significantly to the potential as seen by any individual eigenstate particle. However in a regular eigenstructure array with equal numbers of alternating positive and negative charges, it may be shown that, from a classical perspective [4], the electromagnetic component of the potential felt by an individual charge will, for a large central mass and low particle densities like those encountered in some interesting astrophysical situations be negligible compared to the gravitational potential. For example, consider the charged eigenstate particle in (1) to have charge $q$ and be embedded in a uniform array of other positive and negative charges $q_j(=\pm q)$, then the electrical contribution $V_q$ to the potential energy term from these other charges maybe estimated from the analogous solid-state formula $V_q = 1/4\pi\varepsilon_0 \sum_j \left( q\,q_j / r_j \right) = \alpha q^2 / 4\pi\varepsilon_0 s$ where $r_j$ is the distance to each of the other alternating charges, $\alpha$ is the Madelung constant and $s$ is an average separation distance (analogous to a lattice constant). For typical present-day galactic halo conditions where the average occupied proton/electron particle eigenstate density is $\sim 10^6\,\mathrm{m}^{-3}$ and the central mass $\sim 10^{42}$ kg, the above formula gives the electrostatic potential energy of a typical proton as $\sim 10^{10}$ times smaller than its gravitational potential energy. Limits on the assumption of a neutral central mass $M$ may be obtained by comparing the gravitational potential energy term in (1) with an electrical potential energy term $V_q$ when the central/enclosed mass $M$ has a net charge $Q$, for a proton in a typical eigenstate. In this case these terms are $GMm/r$ and $Qq/4\pi\varepsilon_0 r$ respectively. Substituting typical halo values of $M \sim 10^{42}$ kg and $r \sim 10^{21}$ m, shows that the central mass $M$ would need to have a net charge $Q \sim 10^{14}$ C for the electrical potential energy term to be of comparable magnitude to the gravitational energy term. The existence of such large charge displacements within the galaxy would seem unlikely, but this comparison provides a figure for determining a limit on the validity of the neutrality assumption in terms of the overall galactic charge displacements in the present treatment.

In the present work therefore we ignore any electrical contribution from the charged eigenstate particles themselves to the overall potential as seen by any one particle and also take the central mass $M$ as neutral, subject to the limitations on this as expressed above. Furthermore as regards to eigenstate particles' global contribution to the gravitational potential, since we will be looking predominantly at high $n$, $l$ states where the thickness of their shell-like radial probability distributions is relatively small, the total effect on the potential of the other eigenstate particles may, in this case, be incorporated into an effective net central gravitational potential, analogous to that used in the Hartree-Fock approximation in atomic physics. The condition that this net central potential is weak over the physical extent of the eigenstate is given by the general condition for weak gravity $\varepsilon \equiv GM/rc^2 \ll 1$. For particles in the region of a typical galactic halo for example, $M \sim 10^{42}$ kg and $r \sim 10^{21}$ m, giving $\varepsilon \sim 10^{-6}$, so that this condition is well-satisfied. Since halo density scales as $1/r^2$, $M$ scales directly with $r$ and this condition is satisfied at lower $r$ as well, until for $r \sim 10^{14}$ m, and typical central galactic black hole mass of $10^{38}$ kg, $\varepsilon \sim 10^{-3}$. Thus for the low-$p$ ($< 10^{10}$), high-$n$ ($> 10^{31}$) eigenstates considered here, whose radial extents are $< 10^{10}$ m, and whose radial positions begin at an $r$ at least as large as $r \sim 10^{16} (\gg 10^{14})$, the condition of weak gravity, the $1/r$ approximation to the variation of potential, and the neutrality condition discussed above, are simultaneously satisfied.

The state-to-state transition probability for radiative dipole decay $A_{i,f}$ ($\equiv$ Einstein $A$ coefficient)[1], is analogous to the atomic case. For a transition $n_i \rightarrow n_f$ from initial state $|i\rangle$ to final state $|f\rangle$ this is [10]

$$A_{i,f} = \frac{\omega_{if}^{3} \left| \left\langle f \left| e\mathbf{r} \right| i \right\rangle \right|^2}{3\varepsilon_0 \pi \hbar c^3} = \frac{\omega_{if}^{3} \Pi_{if}^{2}}{3\varepsilon_0 \pi \hbar c^3} \tag{7}$$

where $e$ is the electronic charge, $\varepsilon_0$ the electrical permittivity of free space, $\left| \left\langle f \left| e\mathbf{r} \right| i \right\rangle \right| = \Pi_{if}$ the absolute value of the dipole matrix element for spontaneous decay for the transition $i$ to $f$, $\omega_{if} = \left( \mu G^2 m_p^2 M^2 \big/ \left( 2\hbar^2 \right) \right)\left( 1 \big/ n_f^2 - 1 \big/ n_i^2 \right)$ the angular frequency corresponding to the transition $i$ to $f$, $\mu$ the reduced mass, and the other symbols have their normal meanings. For a state $i$ with multiple decay channels, the reciprocal state lifetime $1/\tau_i$ is then the decay rate sum over all possible decay channels $k$: $1/\tau_i = A_i = \sum_k A_{i,k}$ .

In calculating $1/\tau_i$ there are two problems: the number of transitions $k$ that must be summed over may be extensive for some states, and it may be very difficult to calculate $\Pi_{if}$ ($\equiv \Pi_{ik}$) for the particular $|i\rangle$ and $|k\rangle$ involved in each of the individual $A_{i,k}$. ($\omega_{ik}$ may be calculated from the eigenvalues in a straight forward manner however.) Explicitly $\Pi_{if}$ is written as:

$$\Pi_{if} = \left| \int_0^\infty \int_0^\pi \int_0^{2\pi} R_{n_f,l_f}^* \, Y_{l_f,m_f}^* \, e\mathbf{r} \, R_{n_i,l_i} \, Y_{l_i,m_i} \, r^2 \sin(\theta) \, d\phi \, d\theta \, dr \right|$$
$$= \sqrt{\left( \Pi_{ifx}^2 + \Pi_{ify}^2 + \Pi_{ifz}^2 \right)} \tag{8}$$

where $\Pi_{ifx}$, $\Pi_{ify}$ and $\Pi_{ifz}$ are the $x$, $y$ and $z$ components of the vector inside the modulus in (8). Dipole radiative decay occurs via transitions involving $\Delta m = 0$ (implying $\Pi_{ifx} = \Pi_{ify} = 0$) or $\Delta m = \pm 1$ (implying $\Pi_{ifz} = 0$) and $\Delta l = \pm 1$ (implying transitions must take place between adjacent $l$ columns in Fig. 1.

---

[1] Higher order decays show similar trends to that of dipole decay but rates are several orders of magnitude smaller for each corresponding unit increase in multi-pole order.

$\Pi_{ifx}$, $\Pi_{ify}$ and $\Pi_{ifz}$ may be explicitly written as

$$\Pi_{ifx} = \int_0^\infty \int_0^\pi \int_0^{2\pi} R_{nf,lf}^* \, Y_{lf,mf}^* \, e \, r \, R_{ni,li} \, Y_{li,mi} \, r^2 \sin(\theta) \cos(\phi) \sin(\theta) d\phi d\theta \, dr$$

$$= e \int_0^\infty R_{nf,lf}^* \, r^3 \, R_{ni,li} \, dr \int_0^\pi \int_0^{2\pi} Y_{lf,mf}^* \, Y_{li,mi} \sin(\theta) \cos(\phi) \sin(\theta) d\phi d\theta$$

$$\equiv e \, I_R \, I_{\theta\phi x}$$

(9)

$$\Pi_{ify} = \int_0^\infty \int_0^\pi \int_0^{2\pi} R_{nf,lf}^* \, Y_{lf,mf}^* \, e \, r \, R_{ni,li} \, Y_{li,mi} \, r^2 \sin(\theta) \sin(\phi) \sin(\theta) d\phi d\theta \, dr$$

$$= e \int_0^\infty R_{nf,lf}^* \, r^3 \, R_{ni,li} \, dr \int_0^\pi \int_0^{2\pi} Y_{lf,mf}^* \, Y_{li,mi} \sin(\theta) \sin(\phi) \sin(\theta) d\phi d\theta$$

$$\equiv e \, I_R \, I_{\theta\phi y}$$

(10)

$$\Pi_{ifz} = \int_0^\infty \int_0^\pi \int_0^{2\pi} R_{nf,lf}^* \, Y_{lf,mf}^* \, e \, r \, R_{ni,li} \, Y_{li,mi} \, r^2 \cos(\theta) \sin(\theta) d\phi d\theta \, dr$$

$$= e \int_0^\infty R_{nf,lf}^* \, r^3 \, R_{ni,li} \, dr \int_0^\pi \int_0^{2\pi} Y_{lf,mf}^* \, Y_{li,mi} \cos(\theta) \sin(\theta) d\phi d\theta$$

$$\equiv e \, I_R \, I_{\theta\phi z}$$

(11)

where we have further split the integrals into their radial ( $I_R \equiv \int_0^\infty R_{nf,lf}^*(r) \, R_{ni,li}(r) \, r^3 \, dr$ ) and angular ( $I_{\theta\phi x} \equiv \int_0^\pi \int_0^{2\pi} Y_{lf,mf}^* \, Y_{li,mi} \sin(\theta) \cos(\phi) \sin(\theta) d\theta d\phi$ , etc.) components.

Explicit formulae for the radial and angular integrals in (9), (10) and (11) may be then obtained using (4) and (6) respectively.

The explicit forms of the angular integrals $I_{\theta\phi x}$ , $I_{\theta\phi y}$ and $I_{\theta\phi z}$ vary depending on whether $\Delta m$ is +1, -1 or 0 and whether $\Delta l$ is +1 or -1. We briefly show here the technique for derivation of the case where the initial state is $Y_{li,mi} = Y_{l,m}$ and final state is $Y_{lf,mf} = Y_{l-1,m}$ , that is the combination $\Delta m = 0$ and $\Delta l = -1$ , derivation of the other combinations following along similar lines. In this case $I_{\theta\phi x}$ and $I_{\theta\phi y}$ are zero and we need only calculate $I_{\theta\phi z}$ , $\int_0^\pi \int_0^{2\pi} Y_{lf,mf}^* Y_{li,mi} \cos(\theta) \sin(\theta) d\phi d\theta$ . Assuming for the moment that $l$ and $m$ are both even, using Rodrigues' formula and the binomial theorem, and adjusting the summation limits appropriately, gives the initial state $Y_{li,mi} = Y_{l,m}$ as

$$Y_{l,m} = (-1)^{(m)} \sqrt{\frac{(2l+1)}{4\pi} \frac{(l-m)!}{(l+m)!}} \, (\sin\theta)^m \, e^{im\phi} \times$$
$$\left\{ \sum_{k_i=0}^{(l-m)/2} \left[ (-1)^{k_i} \frac{(2l-2k_i)!(\cos\theta)^{(l-2k_i-m)}}{2^l (l-2k_i-m)!(l-k_i)! k_i!} \right] \right\} \qquad (12)$$

and the final state $Y_{lf,mf} = Y_{l-1,m}$ as

$$Y_{l-1,m} = (-1)^{(m)} \sqrt{\frac{(2l-1)}{4\pi} \frac{(l-m-1)!}{(l+m-1)!}} \, (\sin\theta)^m \, e^{im\phi} \times$$
$$\left\{ \sum_{k_f=0}^{(l-m-1)/2} \left[ (-1)^{k_f} \frac{(2l-2k_f-2)!(\cos\theta)^{(l-2k_f-m-1)}}{2^{l-1}(l-2k_f-m-1)!(l-k_f-1)! k_f!} \right] \right\} \qquad (13)$$

where it is understood that the summation variables extend only to floor integer values for limits which become half integer.

$I_{\theta\phi z}$ then becomes

$$\int_0^{2\pi}\int_0^{\pi} \left( \begin{array}{l} \sqrt{\frac{(2l+1)}{4\pi} \frac{(l-m)!}{(l+m)!}} \, (\sin\theta)^m \, e^{im\phi} \times \\[6pt] \sqrt{\frac{(2l-1)}{4\pi} \frac{(l-m-1)!}{(l+m-1)!}} \, (\sin\theta)^m \, e^{-im\phi} \times \\[6pt] \left\{ \sum_{k_i=0}^{(l-m)/2} \left[ (-1)^{k_i} \frac{(2l-2k_i)!(\cos\theta)^{(l-2k_i-m)}}{2^l (l-2k_i-m)!(l-k_i)! k_i!} \right] \right\} \times \\[6pt] \left\{ \sum_{k_f=0}^{(l-m-1)/2} \left[ (-1)^{k_f} \frac{(2l-2k_f-2)!(\cos\theta)^{(l-2k_f-m-1)}}{2^{l-1}(l-2k_f-m-1)!(l-k_f-1)! k_f!} \right] \right\} \times \\[6pt] \cos\theta\sin\theta \end{array} \right) d\theta \, d\phi \quad (14)$$

The product of the summations may be then combined into a double summation, the powers of $\sin\theta$ and $\cos\theta$ collected, the integrals brought inside the summation and the result simplified to give

$$\left\{ \frac{1}{4\pi} \sqrt{\frac{(2l+1)(2l-1)(l-m)!(l-m-1)!}{(l+m)!(l+m-1)!}} \times \right.$$

$$\left. \left\{ \sum_{k_f=0}^{(l-m-1)/2} \left( \sum_{k_i=0}^{(l-m)/2} \left( \left[ \frac{(-1)^{k_i+k_f}(2l-2k_i)!}{2^{2l-1}k_i!k_f!(l-k_i)!} \times \right. \right. \right. \right. \right. $$

$$\left. \left. \left. \left. \left. \frac{(2l-2k_f-2)!}{(l-2k_f-m-1)!(l-2k_i-m)!(l-k_f-1)!} \times \right] \right) \right) \right\} \right\}$$

$$\int_0^{2\pi}\int_0^{\pi}(\sin\theta)^{2m+1}(\cos\theta)^{(2l-2k_i-2k_f-2m)}\,d\theta\,d\phi \qquad (15)$$

The integral over $\phi$ is just $2\pi$ and the integrals over $\theta$ are of the form $\int_0^{\pi}(\sin\theta)^{2r+1}(\cos\theta)^{2s}\,d\theta$ with $r$ and $s$ are arbitrary positive integers. These may be evaluated using the recurrence relation

$$\int_0^{\pi}(\sin\theta)^{2r+1}(\cos\theta)^{2s}\,d\theta = \frac{2s-1}{2r+2s+1}\int_0^{\pi}(\sin\theta)^{2r+1}(\cos\theta)^{2s-2}\,d\theta$$

and the fact that

$$\int_0^{\pi}(\sin\theta)^{2r+1}\,d\theta = -\frac{\pi^{3/2}\csc(\pi r)}{\Gamma(-r)\Gamma(\tfrac{3}{2}+r)} = \frac{\sqrt{\pi}\,\Gamma(1+r)}{\Gamma(\tfrac{3}{2}+r)} \text{ (for positive integer } r)$$

to give

$$\int_0^{\pi}(\sin\theta)^{2r+1}(\cos\theta)^{2s}\,d\theta = \left(\frac{2s-1}{2r+2s+1}\right)\left(\frac{2s-3}{2r+2s-1}\right)\cdots\left(\frac{1}{2r+3}\right)\int_0^{\pi}(\sin\theta)^{2r+1}\,d\theta$$

$$= \frac{2^{2r}(2s)!\,r!(2r+1)!(r+s)!}{(\tfrac{1}{2}+r)(2r)!\,s!(2r+2s+1)!} \qquad (16)$$

We set $r=m$ and $s=l-k_i-k_f-m$ in (16), substitute these into each of the expanded and collected product sum terms $(\sin\theta)^{2m+1}(\cos\theta)^{2l-2k_i-2k_f-2m}$ of (15).

This yields a final expression for $I_{\theta\phi z}$, so that the angular integrals for this case may be finally reduced to

$$I_{\theta\phi x} = 0$$

$$I_{\theta\phi y} = 0$$

$$I_{\theta\phi z} =$$

$$
\left(
\begin{array}{l}
2^{2m-2l+1} m! \sqrt{\dfrac{(2l+1)(2l-1)(l-m)!(l-m-1)!}{(l+m)!(l+m-1)!}} \; \times \\[4mm]
\left\{
\displaystyle\sum_{k_f=0}^{(l-m-1)/2} \left(
\displaystyle\sum_{k_i=0}^{(l-m)/2} \left[
\begin{array}{l}
\dfrac{(-1)^{k_i+k_f}(2l-2k_i)!(2l-2k_f-2)!(l-k_i-k_f)!}{k_i! \, k_f! \, (l-k_f-1)!(l-k_i)!(2l-2k_f+1)!} \; \times \\[4mm]
\dfrac{(2l-2m-2k_i-2k_f)!}{(l-2k_f-m-1)!(l-2k_i-m)!(l-k_i-k_f-m)!}
\end{array}
\right] \right)
\right)
\right\}
\end{array}
\right) \quad (17)
$$

Similar alternative expressions can be obtained when the quantum numbers $l$ and $m$ are odd or combinations of odd and even, or when the final state is $l+1$ rather than $l-1$. Likewise expressions for $I_{\theta\phi x}$ and $I_{\theta\phi y}$ can be obtained for the cases where $\Delta m = \pm 1$ and in this way explicit expressions for all dipole transition angular integral components may be obtained.

The correctness and exactness of (17) can be verified by comparing cases where the values of $l$ and $m$ are small enough to also allow alternative calculation by direct integration. For example for $l=30$ $m=6$, both integration and (17) yield the exact result of $12\sqrt{6/3599}$.

The derivation of the radial integral, $I_R \equiv \int_0^\infty R_{nf,lf}^*(r)\, R_{ni,li}(r)\, r^3\, dr$ is somewhat more involved but, unlike the angular integrals, only takes one of two possible forms depending on the angular momentum $l+1$ or $l-1$ of the final state. We present here the latter case taking the initial state as $(n_i, l_i) \equiv (n_i, l_i = n_i - p)$ where $p = n_i - l_i$ as defined earlier, and final state as $(n_f, l_f) \equiv (n_f, l_f = l_i - 1 = n_i - p - 1)$. The initial and final states written in terms of $p$ then respectively become, using (4) and (5)

$$R_{n_i,l} = \left( \left( \frac{2}{n_i b_0} \right)^3 \left( \frac{(p-1)!(2n_i-p)!}{2n_i} \right) \right)^{\frac{1}{2}} \exp\left( -\frac{r}{n_i b_0} \right) \left( \frac{2r}{n_i b_0} \right)^{(n_i-p)} \times$$

$$\left( \sum_{k_i=0}^{(p-1)} \frac{(-1)^{k_i} \left( \frac{2r}{n_i b_0} \right)^{k_i}}{(p-k_i-1)!(2n_i-2p+k_i+1)!k_i!} \right)$$

and

$$R_{n_f,l-1} = \left( \left( \frac{2}{n_f b_0} \right)^3 \left( \frac{(n_f-n_i+p)!(n_i+n_f-p-1)!}{2n_f} \right) \right)^{\frac{1}{2}} \times$$

$$\left( \sum_{k_f=0}^{n_f-n_i+p} \frac{(-1)^{k_f} \left( \frac{2r}{n_f b_0} \right)^{k_f}}{(n_f-n_i+p-k_f)!(2n_i-2p+k_f-1)!k_f!} \right) \times$$

$$\exp\left( -\frac{r}{n_f b_0} \right) \left( \frac{2r}{n_f b_0} \right)^{(n_i-p-1)} \tag{18}$$

$I_R$ can be then written as

$$I_R =$$

$$C \times \sum_{k_f=0}^{n_f+p-n_i} \left( \sum_{k_i=0}^{p-1} \left( \frac{\left( -\frac{2}{b_0} \right)^{k_i+k_f} \int_0^\infty \exp\left[ -\frac{n_i+n_f}{n_i n_f b_0} r \right] r^{2n_i-2p+k_i+k_f+2} \, dr}{n_i^{k_i} n_f^{k_f} k_i! k_f! (p-k_i-1)!(n_f+p-n_i-k_f)!} \times \frac{1}{(2n_i-2p+k_f-1)!(2n_i-2p+k_i+1)!} \right) \right) \tag{19}$$

where

$$C = \left(\frac{2}{b_0}\right)^{2n_i - 2p + 2} \frac{\left((p-1)!\,(2n_i - p)!\left(n_f + p - n_i\right)!\left(n_i + n_f - p - 1\right)!\right)^{\frac{1}{2}}}{2 n_i^{\,n_i - p + 2} n_f^{\,n_i - p + 1}}$$

Since

$$\int_0^\infty \exp\left[-\frac{n_i + n_f}{n_i n_f b_0}\,r\right] r^{2n_i - 2p + k_i + k_f + 2}\,dr$$

$$= \left(2(n_i - p) + k_i + k_f + 2\right)!\left(\frac{b_0\,n_i\,n_f}{n_i + n_f}\right)^{\left(k_i + k_f + 2n_i - 2p + 3\right)}$$

the expression for $I_R \left(\equiv \int_0^\infty R_{nf,lf}^*(r)\,R_{ni,li}(r)\,r^3\,dr\right)$ can therefore finally be written as

$$I_R = \frac{2^{2n_i - 2p + 1} n_i n_f^{\,2} b_0}{\left(n_i + n_f\right)^3}\left(\frac{n_i n_f}{\left(n_i + n_f\right)^2}\right)^{n_i - p} \times$$

$$\left((p-1)!\,(2n_i - p)!\left(n_f + p - n_i\right)!\left(n_i + n_f - p - 1\right)!\right)^{\frac{1}{2}} \times \qquad (20)$$

$$\sum_{k_f = 0}^{n_f + p - n_i}\left(\sum_{k_i = 0}^{p-1}\left(\frac{\left(-2n_i\right)^{k_f}\left(-2n_f\right)^{k_i}}{\left(n_i + n_f\right)^{k_i + k_f}\left(p - k_i - 1\right)!\,k_i!\left(n_f + p - n_i - k_f\right)!\,k_f!} \times \frac{(2n_i - 2p + k_i + k_f + 2)!}{(2n_i - 2p + k_f - 1)!(2n_i - 2p + k_i + 1)!}\right)\right)$$

Again the exactness of (20) may be checked by substituting low values of $n_i$, $n_f$ and $p$ directly and comparing these with results obtained by direct integration of $\int_0^\infty R_{nf,lf}^*(r)\,R_{ni,li}(r)\,r^3\,dr$. For example, if $n_i = 6$, $n_f = 3$, $p = 4$ (so that $l_i = 2$ and $l_f = 1$) both expressions give a value of $4096\sqrt{70}\,b_0\,/\,19683$.

Combining (7), (17) and (20) then enables any state-to-state transition rate to be calculated for $\Delta l = -1$ and $\Delta m = 0$. As a check on the validity of these expressions, they may be easily converted to their equivalent atomic counterparts. For example the decay rate for the $n_i = 2, l_i = 1 \rightarrow n_f = 1, l_f = 0$ transition in hydrogen may be calculated using the present treatment by substituting parameter values relevant to this atomic transition yielding $6.2 \times 10^8 \, \text{s}^{-1}$ compared to literature values [10, 11] of $6.3 \times 10^8 \, \text{s}^{-1}$. Decay of the "Rydberg" atomic hydrogen $n_i = 10$, $l_i = 0$ and $n_i = 10$, $l_i = 1$ states are more complicated because they involve addition of rates over a greater number of decay channels and also transitions involving other $\Delta l$ and $\Delta m$ values ( $\Delta l = \pm 1$ and $\Delta m = 0, \pm 1$ ). Using equations (7), (17) and (20) and other equivalent expressions the present treatment yields lifetimes for the hydrogen Rydberg states $n_i = 10$, $l_i = 0$ and $n_i = 10$, $l_i = 1$ of $2.0$ and $0.19 \, \mu s$ compared with the values obtained from [12] of $2.1$ and $0.21 \, \mu s$ respectively.

## 3. Dipole matrix element and transition rate approximations for transitions where at least one state has *p* = 1

The interesting states are those that have very low decay rates and that are weakly interacting as these would be the ones most likely to exist naturally. We are concerned therefore to know upper limits on the decay and interaction rates of particular types of transitions. In the case of dipole decay these depend on the values of $\Pi_{if}$ and $\omega_{if}$ in (7). The value of $I_R$ is critical in determining the decay rate because it depends on $r$ that can become very large. The smallness of angular components of $\Pi_{if}$ therefore becomes an important consideration in limiting the transition rate only if $I_R$ is not small. Consequently it is the value of $I_R$ that we wish to primarily investigate here, because of its dramatic effect on $\Pi_{if}$ and on the resulting transition rates.

No approximations have been made in the derivation of $I_R$ in (16) and (20) so that these equations enable the possibility of obtaining tractable expressions for the radial component of the dipole matrix elements corresponding to any dipole transitions within the conditions set out earlier in this paper. For example substituting $n_f = n_i - 1$ and $p = 1$ into (20) gives an explicit expression for the radial component of transitions of the type B to A ( $(n_i = n_i, l = n_i - 1) \rightarrow (n_f = n_i - 1, l = n_i - 2)$ ) shown in figure 1 as:

$$\int_0^\infty R_{nf,lf}^*(r)\, r^3\, R_{ni,li}(r)\, dr = 2^{2n_i} b_0 \left( \frac{n_i(n_i - 1)}{(2n_i - 1)^2} \right)^{n_i + 1} \sqrt{(2n_i - 1)^3 (2n_i - 2)} \quad (21)$$

which, for large $n_i$ becomes $\approx b_0 n_i^2$. Again the agreement of the formulae for these low-$p$-to-low-$p$ state transitions may be verified when the values of $n$, $l$ and $p$ are sufficiently low for $I_R$ to be manageable by direct integration of the polynomial representations.

A similar dependence on $b_0 n_i^2$ exists for adjacent state to state transitions on the $p = 2$ diagonal and hence yields radiative decay rates essentially the same as that for transitions of the type B to A. It can be shown using (20) that whenever transitions take place between two '$p$-turning point' Laguerre eigenfunctions, that is transitions that take place along the same $p$-diagonal so that $n_i \to n_i - 1$ and $l_i \to l_i - 1$, then the decay rate is $b_0 n_i^2$ provided $n_i \gg p$.

Transitions like C to A of figure 1 involve overlap integrals where the radial component wave functions have very different shapes (A is a 1-turning point function while C is a 2-turning point function). As a result, it would be expected in this case that the radial part of the overlap integral would be much smaller than the B to A type transitions. This is indeed the case and (20) gives

$$\int_0^\infty R_{nf,lf}^*(r)\, r^3\, R_{ni,li}(r)\, dr =$$
$$2^{2n_i - 2} b_0 \left( \frac{n_i(n_i - 2)}{(2n_i - 2)^2} \right)^{n_i} \sqrt{(2n_i - 2)(2n_i - 3)(2n_i - 4)} \quad (22)$$

which reduces to the approximate value of $b_0 n_i^{\frac{3}{2}} / \sqrt{2}$ again provided $n_i \gg p$.

A general formula for the value of $I_R$ may be obtained for a more general transition such as D to A of figure 1, which originates from an arbitrary $p$-turning point radial Laguerre polynomial state D $(n_i, l_i = n_i - p)$ and ends on a 1-turning point state like A $(n_f = n_i - p, l_f = n_f - 1$, lying on the left diagonal). In general for such transitions it is possible for $p$ to be much larger than 1 (but still much less than $n_i$). It may be shown that in this case (20) reduces to a single summation

$$I_R = \frac{2^{2n_i-2p+1} n_i n_f{}^2 b_0}{\left(n_i+n_f\right)^3} \left(\frac{n_i n_f}{\left(n_i+n_f\right)^2}\right)^{n_i-p} \times$$

$$\left((p-1)!(2n_i-p)!\left(n_f+p-n_i\right)!\left(n_i+n_f-p-1\right)!\right)^{\frac{1}{2}} \times$$

$$\sum_{k_i=0}^{p-1}\left(\frac{(2n_i-2p+k_i+2)!\left(-2n_f\right)^{k_i}}{\left(n_i+n_f\right)^{k_i}(p-k_i-1)!(2n_i-2p+k_i+1)!k_i!(n_f+p-n_i)!(2n_i-2p-1)!}\right)$$

(23)

which may be written as

$$\frac{b_0}{2}\sqrt{(p-1)!}\left(\frac{n_i n_f}{\left(n_i+n_f\right)^2}\right)^{(n_f+1)} 2^{2n_i-1}\left(\prod_{i=0}^{p}\left(n_i+n_f-i\right)^{\frac{1}{2}}\right) \times$$

$$\left(\sum_{i=1}^{p}\left((-1)^{i-1}\left(\frac{n_f}{\left(n_i+n_f\right)}\right)^i \frac{2^{i-2p+2}\left(2n_f+i+1\right)}{(i-1)!(p-i)!}\right)\right)$$

(24)

Carrying out the summation over *i* and simplifying the product, (24) becomes

$$b_0\left(\frac{4n_i n_f}{\left(n_i+n_f\right)^2}\right)^{(n_f+1)} \frac{p^{p-2} n_i n_f}{\left(n_i+n_f\right)^p}\sqrt{\frac{\left(n_i+n_f\right)!}{(p-1)!\left(2n_f-1\right)!}}$$

(25)

which is so far an exact expression.

Using Stirling's formula one excellent working approximation to (25) even when *p* is relatively small is

$$\frac{b_0\, p^{p/2} n_f{}^{9/4}\left(2n_f+2p\right)^{n_f+2}\left(2n_f+p+1\right)^{n_f+p/2+1/4}}{\left(e^2\pi p^7\right)^{1/4}\left(2n_f+p\right)^{2+2n_f+p}}$$

(or often more usefully written in its log form

$$\exp\left(\begin{array}{l} \log\left(\dfrac{b_0\, p^{p/2} n_f^{\,9/4} \left(2n_f + 2p\right)^2 \left(2n_f + p + 1\right)^{p/2 + 1/4}}{\left(e^2 \pi p^7\right)^{1/4} \left(2n_f + p\right)^{2+p}}\right) + \\[2em] n_f \log\left(\dfrac{\left(2n_f + 2p\right)\left(2n_f + p + 1\right)}{\left(2n_f + p\right)^2}\right) \end{array}\right) \qquad (26)$$

to avoid terms of order $n_f^{\,n_f}$).

A simpler but more approximate expression can be used for calculations that involve large values of $n_i$ and $p$, provided $n_i \gg p \gg 1$. Then $n_i \approx n_f \,(= n_i - p)$, and using $\lim\limits_{n \to \infty} \left(\left(2n - p + 1\right)/\left(2n - 2p\right)\right)^n = \exp\left(\left(p + 1\right)/2\right)$, enables (26) to be simplified to

$$I_R = b_0 n_i \left(\frac{2}{\pi p}\right)^{\frac{1}{4}} \left(\frac{e}{2}\right)^{\frac{p}{2}} \left(\frac{p}{n_i}\right)^{\frac{p-3}{2}} \qquad (27)$$

The usefulness of the approximate equations (26) and (27) in determining $I_R = \int_0^\infty R_{nf.lf}^*(r)\, r^3 R_{ni,li}(r)\, dr$ may be tested by comparison with the direct integral using lower values for $n_i$ and $p$ and this is shown in table 1 in units of $b_0$.

**Table 1.** Value of $I_R = \int_0^\infty R_{nf.lf}^*(r)\, r^3 R_{ni,li}(r)\, dr$ obtained by (a) direct integration, (b) using equation (26), and (c) using equation (27), for an initial state $n_i$, $p_i$ and final state ending on a $p_f = 1$ diagonal of figure 1.

| $n_i$ | $p$ | Radial overlap integral | Value using equation (26) | Value using equation (27) |
|-------|-----|-------------------------|---------------------------|---------------------------|
| $10^2$ | 5 | 5.76 | 5.80 | 6.43 |
| $10^2$ | 10 | $5.58 \times 10^{-2}$ | $5.60 \times 10^{-2}$ | $7.37 \times 10^{-2}$ |
| $10^2$ | 50 | $6.88 \times 10^{-5}$ | $6.89 \times 10^{-5}$ | $6.08 \times 10^{-3}$ |
| $3 \times 10^4$ | 5 | 6.38 | 6.43 | 6.43 |
| $3 \times 10^4$ | 10 | $4.70 \times 10^{-8}$ | $4.72 \times 10^{-8}$ | $4.73 \times 10^{-8}$ |
| $3 \times 10^4$ | 50 | $1.10 \times 10^{-58}$ | $1.10 \times 10^{-58}$ | $1.12 \times 10^{-58}$ |

The approximations (26) and (27) highlights the rapidity with which $I_R = \int_0^\infty R_{nf,lf}^*(r) \, r^3 R_{ni,li}(r) \, dr$ approaches zero as $p$ increases, because of the factor $\left( \dfrac{p}{n_i} \right)^{\frac{p-3}{2}}$. Table 2 provides values of $I_R$ for a range of values of $p$ and $n_i$ illustrating this effect. The overwhelming smallness of $I_R$ and the resulting overwhelming smallness of the dipole matrix element $\Pi_{if}$ means that state-to-state decay rates which start on any large-$p$ diagonals and end on the $p = 1$ diagonal (or as it turns out on any low-$p$ diagonal) are negligibly small, even despite sizeable values of $n_i - n_f$ and the cubic dependence of $A_{i,f}$ on $\omega_{if}$. This is an extremely interesting result since it means that all the high-$n$, low-$p$ eigenstates in gravitational wells discussed here all exhibit the unusual property of having extremely long lifetimes. As a consequence they will, despite their high-valued quantum parameters, also exhibit inherent long-term stability and an inability to coalesce by radiative decay or interact strongly with other electromagnetic radiation.

**Table 2.** Value of $I_R = \int_0^\infty R_{nf,lf}^*(r) \, r^3 R_{ni,li}(r) \, dr$ obtained using equation (26), for an initial 'deep' state D (quantum parameters $n_i$, $p_i \gg 1$) that ends on the $p = 1$ diagonal of figure 1.

| $n_i$ | $p_i$ | $\int_0^\infty R_{nf,lf}^*(r) \, r^3 R_{ni,li}(r) \, dr$ (in units of $b_0$) |
|---|---|---|
| 1000 | 1 | $\sim 10^6$ |
| 1000 | 5 | $\sim 6$ |
| 1000 | 20 | $\sim 3 \times 10^{-11}$ |
| 1000 | 100 | $\sim 10^{-40}$ |
| $10^{30}$ | 1 | $\sim 10^{60}$ |
| $10^{30}$ | 5 | $\sim 6$ |
| $10^{30}$ | 10 | $\sim 10^{-71}$ |
| $10^{30}$ | $10^{20}$ | $\sim 10^{-5 \times 10^{20}}$ |
| $10^{30}$ | $10^{26}$ | $\sim 10^{-2 \times 10^{26}}$ |
| $8 \times 10^{33}$ | $5 \times 10^{31}$ | $\sim 10^{-5 \times 10^{31}}$ |

## 4. Conclusion

Exact general expressions have been determined for the transition probability per unit time of the electromagnetic decay of charged particles in gravitational eigenstates for any dipole-allowed transitions in weak gravity. Additionally tractable approximations have been obtained for certain specific transitions involving the interesting set of high-$n$, low-$p$ eigenstates, including a general one for transitions ending on the $p = 1$ diagonal. The generality of these types of formulae also give them application in the study of other large-$n$ quantum systems such as Rydberg states in atomic systems. The most significant conclusion to come out of this present study is that the very high-$n$, low-$p$ gravitational eigenstates have extremely long lifetimes (potentially many times the age of the universe in the situations considered here).

The speculation was raised in the introduction that gravitational eigenstates might exist naturally elsewhere in the universe. Small structures in low-central-mass wells are generally weakly bound (as demonstrated in [1,2]), so that if well bound eigenstructures exist they are likely to be of macroscopic size. This is not a problem in itself, but it is hard to see how these structures could have formed in all but the earliest times in cosmic history since it can be shown [4] that the probability for a localised object of any reasonable mass decaying into an eigenstate in any time is negligible. The formation of such structures would require the existence of strong gravitational potential wells at a time in universal history when the global particle density over the well region was sufficiently high and the temperature sufficiently low to enable decay into eigenstates, analogous to plasma recombination. Indeed if such conditions were met it would be expected that eigenstates would form and ultimately populate the longest lived low-$p$ levels. It turns out that such conditions are possible at certain critical times in universal history and the formation mechanisms associated with these structures will be the subject of a later paper. It is clear however that the resulting stability of particles occupying eigenstates would mean that if such a structure formed, it would be stable and have an inability to gravitationally collapse. It is also expected that it would not intrinsically emit or scatter radiation to any significant degree and therefore be essentially invisible with respect to external electromagnetic radiation [4]. This latter point will be discussed further in a companion paper.

The approximation for calculating the dipole transition rate for those transitions involving $p = 1$ is a useful one, but it is clearly necessary to develop further approximate techniques for the calculation of other dipole transitions with more generalized values of $p$. In this regard, the derivations in the present paper continue to have limitations. In the case of the gravitationally bound systems with quantum parameters $n$, $l \geq 10^{30}$ for example, the direct use of (20) is limited because of the very large factorial functions (despite the use of Stirling's approximation) and also because it involves summations with unrealistically large numbers of terms. Different approaches will be needed to deal with these types of transitions, where both the initial and final $p$ values of the states are large.

Other areas requiring more detailed examination are those of multi-pole decay and transitions induced by other types of interactions such as particle collisions. These investigations will be the subject of later papers. It would appear however from this initial work that the high-$n$, low-$p$ gravitational eigenstates are particularly interesting to study theoretically because of their extremely long lifetimes and the implications this has for their expected behaviour and appearance.